# Theory of the lower critical magnetic field for a two-dimensional superconducting film in a non-uniform field

Thomas R. Lemberger and John Draskovic
Dept. of Physics
The Ohio State University
Columbus, OH 43210

We consider the first appearance of vortices in a two-dimensional (2-D) superconducting film exposed to a non-uniform magnetic field, $\boldsymbol{B}_a$, produced by a nearby coil. The film has "infinite" radius, $R_f$, and thickness $t$ about equal to the coherence length, $\xi$. The coil is approximated as a point dipole. We find that the first vortex-bearing state to appear has both a vortex and an antivortex. The Gibbs free energy of this state is lower than the vortex-free state when the maximum applied perpendicular field, i.e., the applied field, $B_0$, at the origin, exceeds the external critical field: $B_{c1}^0 = \frac{4\sqrt{2}\Lambda}{R}\frac{\Phi_0}{4\pi\Lambda^2}ln\left(\frac{\Lambda}{\xi}\right)$, where $\frac{\Phi_0}{4\pi\Lambda^2}ln\left(\frac{\Lambda}{\xi}\right) \equiv B_{c1}^{2D}$ is the intrinsic critical field in 2-D, $\Lambda \equiv 2\lambda^2/t$ is the 2-D penetration depth introduced by Pearl, and $\lambda$ is the bulk penetration depth. The prefactor, $4\sqrt{2}\Lambda/R$, is calculated in the strong-screening regime, $\Lambda/R \ll 1$. $R$ is the radial distance at which the applied perpendicular field, $B_{a,z}(\rho)$, changes sign. In the lab, the onset of vortex effects generally occurs at a field much higher than $B_{c1}^0$, indicating that vortices are inhibited by the vortex-antivortex unbinding barrier, or by pinning.

# I. INTRODUCTION

This paper presents a calculation of the external thermodynamic critical field, $B_{c1}^0$, for an infinite radius, 2-D superconductor in a non-uniform applied magnetic field. An interesting result is that the first vortex-bearing state to appear has both a vortex, near the origin, and an antivortex, far from the origin but not at infinity. Implicit in the calculation is the notion that vortices arise from thermally excited, bound vortex-antivortex (V-aV) pairs that break into independent vortices by overcoming the free-energy barrier that binds them. When they unbind, vortices move toward the center of the film while antivortices move away until the long-range V-aV attraction stops them. In the lab, vortex physics often occurs at a much larger field than we calculate here, so our various simplifications are accurate enough for the purpose of the paper.

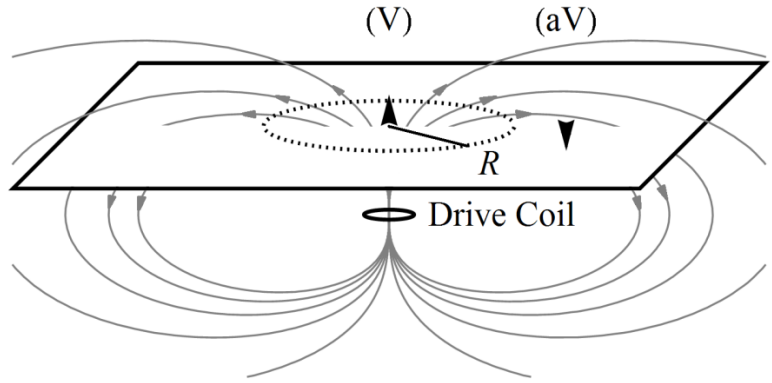

Fig. 1. Diagram illustrating: the film (square), field lines from the drive coil, the drive coil located $R/\sqrt{2}$ below the film, a circle of radius $R$ (dotted) at which $B_{a,z}(\rho)$ changes sign, a vortex at the origin (up arrowhead) and an antivortex (down arrowhead).

This paper is motivated by our interest in knowing the smallest applied magnetic field at which vortices might first appear in a two-coil experiment [e.g., Refs. 1-7] for measuring superfluid density. In a two-coil experiment, a drive coil is located just below the center of a superconducting film, Fig.1. A current in the coil produces a non-uniform magnetic field whose $z$-component, $B_{a,z}(\rho)$, often is largest at the center of the film and reverses sign at a radial distance, $R$, that is much smaller than the radius, $R_f$, of the film. Experimentally, dissipation due to vortices arises when the applied field exceeds a certain well-defined threshold.[5-7] In the following, a "vortex" has its magnetic field parallel to the applied field at the origin; an "antivortex" has the opposite orientation.

This paper complements theoretical work [e.g., Refs. 8-10] on the converse problem of a finite-radius, 2-D superconducting film in the uniform perpendicular magnetic field of a surrounding coil. Applying the macroscopic concept of demagnetization,[11] Fetter and Hohenberg[9] proposed that the "external" lower critical field, $B_{c1}^0$, in this geometry should be the intrinsic 3-D critical field, $B_{c1}^{3D} = (\Phi_0/4\pi\lambda^2)ln(\lambda/\xi)$, ($\lambda$ = penetration depth and $\xi$ = coherence length) reduced by a factor $D \approx \pi t/2R_f$. In a later paper,[10] Fetter used a Ginzburg-Landau approach to obtain a more accurate result for $B_{c1}^0$. This latter theory includes the interaction between a vortex and the perimeter of the film.

Mawatari and Clem[12] (MC) consider vortices created in infinite-radius films by an inhomogeneous applied field, but their films are thick enough to sustain a vortex parallel to the film. They calculate a critical field by assuming that vortices and antivortices first appear when the magnetic field parallel to the film is large enough to create a vortex that arcs into the film. When the middle of the vortex pops through the back of the film, the two ends of the vortex remain in the film, forming a vortex-antivortex pair. We believe that MC calculate the applied field at which the free-energy barrier for creation of a vortex-antivortex pair vanishes, (e.g., a Bean-Livingston type barrier[13]). We find a much lower critical field.

Experimental work on nonlinear effects in two-coil experiments traces back through Claassen and collaborators[5,6] to that of Scharnhorst.[7] The latter found that nonlinear effects appear in quench-condensed Sn and In films when the Meissner screening supercurrent density is near the depairing current density, $J_c(T)$. Since $J_c(T)$ is inversely proportional to $\xi$ and $\Lambda$, this finding offers the possibility that a combination of linear and nonlinear measurements can be used to determine $\xi$ in novel superconducting materials.

## II. CALCULATION

In this section we calculate the Helmholtz free energy of vortices and the work done by the drive coil's current supply when a vortex appears. From these we construct the Gibbs free energy difference, $\Delta G$, between the vortex-free Meissner state and a state with a single V-aV pair. Since the applied field is non-uniform, we define $B_{c1}^0$ as the value of the maximum applied perpendicular field, $B_0$, at the point where $\Delta G < 0$ and the V-aV configuration is stable, i.e., $\Delta G$ is a minimum as a function of the separation between vortex and antivortex.

### A. Applied magnetic field and Meissner screening supercurrent density

The perpendicular component of the magnetic field from a point dipole located $R/\sqrt{2}$ below the film is (Fig. 2):

$$B_{a,z}(\rho) = B_0 \frac{1 - \rho^2/R^2}{(1 + 2\rho^2/R^2)^{5/2}} \tag{1}$$

The maximum applied perpendicular field is $B_0$, at the origin; the maximum applied parallel field is $0.4 B_0$, at $\rho = R/2\sqrt{2}$.

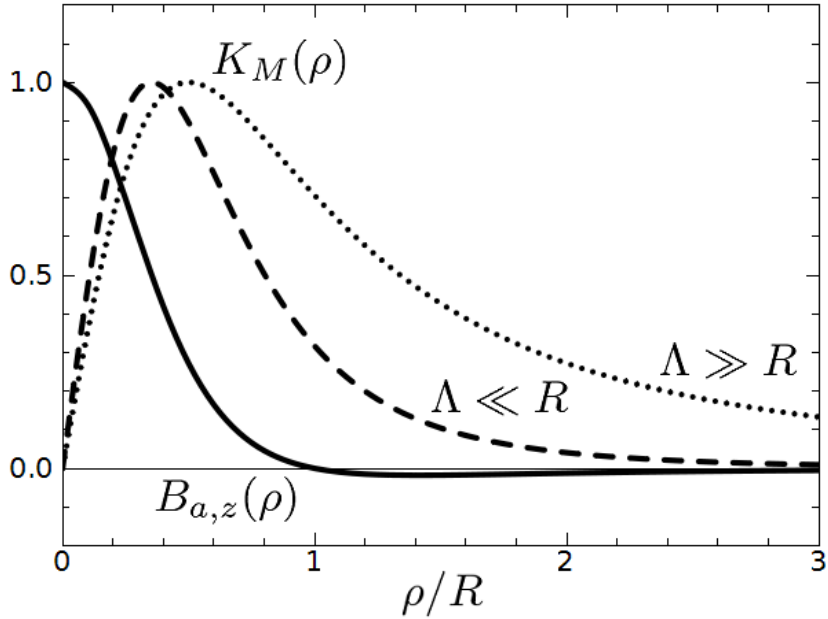

Fig. 2. Applied magnetic field, $B_{a,z}(\rho)/B_0$, (solid curve) from a point dipole placed $R/\sqrt{2}$ below the film, and normalized Meissner supercurrent densities, $|K_M(\rho)|$, for the strong-screening [dashed, Eq. (2)] and weak-screening [dotted, Eq. (3)] limits.

The superconductor responds to the applied field with a Meissner supercurrent density, $J_M(\rho)$, that is uniform through the film thickness when $t \ll \lambda$. The sheet supercurrent density is: $K_M(\rho) = -2A(\rho)/\mu_0 \Lambda$, where $A(\rho)$ is the vector potential and $1/\Lambda \equiv t/2\lambda^2$ is proportional to the areal superfluid density. $\Lambda$ is the 2-D penetration depth identified by Pearl.[8] In the strong-screening regime, $\Lambda/R \ll 1$, $K_M(\rho)$ has the same dependence on $\rho$ as the parallel component of the applied field at the film surface [in square brackets in Eq. (2)]:

$$\mathbf{K}_M(\rho) = -\hat{\boldsymbol{\theta}} \frac{2XR}{\mu_0 \Lambda} \left[ \frac{3B_0}{\sqrt{2}} \frac{\rho/R}{(1+2\rho^2/R^2)^{5/2}} \right]. \quad (\Lambda/R \ll 1) \qquad (2)$$

This result follows from $\nabla \times \mathbf{B} = \mu_0 \mathbf{J}$ and the fact that in the strong-screening regime $\mathbf{B}$ on the backside of the film is much smaller than the applied field. The former implies that $\mathbf{J}$ in the film is proportional to the discontinuity in parallel field, i.e., to the applied parallel field. In the weak-screening regime, the field produced by supercurrents is much smaller than the applied field, so $K_M(\rho)$ is essentially proportional to the vector potential of the point dipole drive coil:

$$\mathbf{K}_M(\rho) = -\hat{\boldsymbol{\theta}} \frac{XB_0 R}{\mu_0 \Lambda} \frac{\rho/R}{(1+2\rho^2/R^2)^{3/2}}. \quad (\Lambda/R \gg 1) \qquad (3)$$

In Eqs. (2) and (3), $XB_0$ is the net field at the center of the film, i.e., the applied field plus the field from screening supercurrents. Self-consistency finds:

$$X = \frac{1}{1+R/\Lambda} \approx \frac{\Lambda}{R}, \quad (\Lambda/R \ll 1) \qquad (4)$$

a result obtained analytically by Gilchrist and Brandt.[14] In the weak-screening regime, $\Lambda/R > 1$, $X$ approaches unity from below: $X \approx 1 - R/2\sqrt{2}\Lambda$.

**B. Free energy of an isolated vortex**

There are several contributions to the free energy of a vortex: kinetic energy, magnetic field energy, and energy of its normal core. The kinetic free energy, *KE*, of vortices comes from integrating the term proportional to $J_S^2$ in the G-L free-energy density:

$$KE = \frac{\mu_0 \lambda^2}{2} \int dV\, J_S^2(\mathbf{r})$$

$$= \frac{\mu_0 \Lambda}{4} \int dA \left[ K_M^2(\boldsymbol{\rho}) + \sum_i K_{V,i}^2 + 2\sum_i \mathbf{K}_{V,i} \cdot \mathbf{K}_M(\boldsymbol{\rho}) + \sum_{i \neq j} \mathbf{K}_{V,i} \cdot \mathbf{K}_{V,j} \right], \quad (5)$$

with the acknowledgement that $\mathbf{K}_S(\boldsymbol{\rho}) = \mathbf{K}_M(\boldsymbol{\rho}) + \sum \mathbf{K}_{V,i}$ is the total supercurrent density at point $\boldsymbol{\rho}$, and $\sum \mathbf{K}_{V,i}$ is the sum of vortex currents at $\boldsymbol{\rho}$. The 2$^{nd}$, 3$^{rd}$, and 4$^{th}$ terms in the integral represent the kinetic energy of isolated vortices, the interaction of vortices with screening supercurrent and with each other.

Not too close to the film perimeter, the sheet supercurrent density, $\mathbf{K}_V(\rho)$, of a vortex at the origin is:

$$\mathbf{K}_V(\rho) \approx \hat{\theta} \frac{\Phi_0}{\pi \mu_0 \Lambda^2} \frac{1}{(\rho/\Lambda)(1+\rho/\Lambda)} \text{ for } \xi < \rho \ll R_f; \ \mathbf{K}_V = 0 \text{ for } \rho < \xi. \quad (6)$$

This approximation is asymptotically correct for small and large $\rho$ ($\xi \ll \rho \ll \Lambda$ and $\Lambda \ll \rho \ll R_f$, respectively), and is within 10% for $\rho \approx \Lambda$.[8,9] $\Phi_0 \equiv h/2e$ is the flux quantum. The corresponding vector potential and magnetic field in the plane of the film are:

$$A_{V,\theta}(\rho) = \frac{\Phi_0}{2\pi\rho} - \frac{\mu_0 \Lambda}{2} K_{V,\theta}(\rho) = \frac{\Phi_0}{2\pi\Lambda} \frac{1}{1+\rho/\Lambda} \quad (7)$$

$$B_{V,z}(\rho, z=0) \approx \frac{\Phi_0}{2\pi\Lambda^2} \frac{1}{(\rho/\Lambda)(1+\rho/\Lambda)^2}. \quad (8)$$

$B_{V,z}(\rho)$ integrates to a net flux through the film of $\Phi_0$, despite its mild divergence as $\rho \to 0$. Within the volume that is at least several $\Lambda$'s away from the film ($\Lambda \ll |z| \ll R_f$) and not too far from the z-axis ($\rho \ll R_f$), the vortex field is that of a magnetic monopole at the origin, e.g.:

$$B_{V,z}(\rho, |z| \gg \Lambda) \approx \frac{\Phi_0 |z|}{2\pi(\rho^2 + z^2)^{3/2}}. \tag{9}$$

$B_{V,z}(\rho, |z| \gg \Lambda)$ integrates to a net flux of $\Phi_0$ through any "plane" parallel to the film, as long as the plane's radius is much greater than |z| but much less than $R_f$, and the plane is centered on the z-axis.

The kinetic free energy of one vortex, $KE_V$, is the second term on the right-hand side of Eq. (5). Using Eq. (6) and integrating yields:

$$KE_V \approx \frac{\mu_0 \Lambda}{4} 2\pi \int_\xi^\infty d\rho \rho K_V^2(\rho) = \frac{\Phi_0^2}{2\pi\mu_0\Lambda}\left[ \ln\left(\frac{\Lambda}{\xi}\right) - 1 \right]. \tag{10}$$

The self-magnetic-field energy of a vortex is: $U_{V,mag} \equiv \int dV\, B_V^2/2\mu_0 = \int dA\, \mathbf{K}_V \cdot \mathbf{A}_V / 2$, where the second equality comes from writing $B_V^2$ as $\mathbf{B}_V \cdot \nabla \times \mathbf{A}_V$ and integrating by parts. Using Eqs. (6) and (7), we find: $U_{V,mag} \approx \Phi_0^2/2\pi\mu_0\Lambda$. The vortex core free energy is: $U_{V,core} \approx \xi^2 t(B_c^2/2\mu_0) = \Phi_0^2/8\pi^2\mu_0\Lambda$, where $B_c = \Phi_0/2\sqrt{2}\pi\lambda\xi$ is the thermodynamic critical field. Thus, the isolated-vortex free energy is:

$$U_V \approx \frac{\Phi_0^2}{2\pi\mu_0\Lambda}\left[ \ln\left(\frac{\Lambda}{\xi}\right) + \frac{1}{4\pi} \right] \approx \frac{\Phi_0^2}{2\pi\mu_0\Lambda} \ln\left(\frac{\Lambda}{\xi}\right). \tag{11}$$

There is considerable uncertainty in the constant, "$1/4\pi$", in Eq. (11). Since $ln(\Lambda/\xi)$ typically ranges from 5 to 8 for very thin films, this constant is usually neglected. If we keep it, then we find in Sec. C the reasonable result that the net energy of a V-aV pair separated by $\xi$ is just the energy of two vortex cores.

### C. Interaction of vortices with screening supercurrent, applied field, and each other

The net interaction of vortices with the Meissner screening supercurrent vanishes due to a cancellation. The kinetic interaction energy is the third term in the integral in Eq. (5). For example, for a vortex at the origin this energy is:

$$U_{VM}(\rho_V = 0) = \frac{\mu_0 \Lambda}{2} \int dA \mathbf{K}_V(\rho) \cdot \mathbf{K}_M(\rho) \approx -\frac{2\Phi_0 B_0 \Lambda}{\mu_0} \quad . \quad (\Lambda/R \ll 1) \quad (12)$$

The overlap field energy is the cross term in the integral, $\int dV (\mathbf{B}_M + \mathbf{B}_V)^2 / 2\mu_0$, where $\mathbf{B}_M(\mathbf{r})$ is the applied field plus the field from the Meissner supercurrent when no vortices are present, and $\mathbf{B}_V(\mathbf{r})$ is the field of a vortex. Writing $\int dV \mathbf{B}_V \cdot \mathbf{B}_M / \mu_0$ as $\int dV \mathbf{B}_V \cdot (\nabla \times \mathbf{A}_M) / \mu_0$, using Maxwell's equation, $\nabla \times \mathbf{B}_V = \mu_0 \mathbf{J}_V$, and integrating by parts yields:

$$U_{ovrlp} \equiv \int dV \mathbf{B}_V \cdot \mathbf{B}_M / \mu_0 = \int dA \mathbf{K}_V \cdot \mathbf{A}_M . \quad (13)$$

Replacing $\mathbf{A}_M$ with $-\mu_0 \Lambda \mathbf{K}_M / 2$ shows that $U_{ovrlp}$ cancels $U_{VM}$.

The interaction energy $V_{12}$ between a vortex and an antivortex separated by $\rho_{12} \equiv |\mathbf{\rho}_2 - \mathbf{\rho}_1|$ has two contributions. One is the fourth term in the integral for kinetic energy, Eq. (5). This term diverges as $ln(\Lambda/\rho_{12})$ for small separation, and it falls of as $1/\rho_{12}^2$ at large separation. The other is the overlap magnetic field energy: $V_{12}^{B \cdot B} \equiv \int dV \mathbf{B}_V(\mathbf{r}) \cdot \mathbf{B}_{aV}(\mathbf{r}) / \mu_0$. Because vortex fields are those of monopoles over a significant volume, the integrand $\mathbf{B}_V(\mathbf{r}) \cdot \mathbf{B}_{aV}(\mathbf{r})$ is significant out to $|z|$ comparable to the spacing between vortices. Thus, $V_{12}^{B \cdot B}$ falls off slowly, as $1/\rho_{12}$, as is seen by evaluation of the equivalent areal integral:

$$V_{12}^{B \cdot B} = \int dA \mathbf{K}_V \cdot \mathbf{A}_{aV} \approx -\frac{\Phi_0^2}{\pi \mu_0 \Lambda} \frac{1}{1 + \rho_{12}/\Lambda} \ . \tag{14}$$

As Pearl[8] first showed, the sum of these terms is:

$$V_{12} \approx -\frac{\Phi_0^2}{\pi \mu_0 \Lambda} \ln\left(1 + \frac{\Lambda}{\rho_{12}}\right), \tag{15}$$

which displays the logarithmic increase of kinetic energy at $\rho_{12} \ll \Lambda$ and the $1/\rho_{12}$ falloff of the field energy at $\rho_{12} \gg \Lambda$.

As mentioned above, Eqs. (11) and (15) show that when a vortex and antivortex are close together, their Helmholtz free energy, $F_{V-aV}(\rho_{12}) = 2U_V + V_{12}(\rho_{12})$, is just that of two vortex cores: $F_{V-aV}(\rho_{12} \approx \xi) \approx \Phi_0^2/4\pi^2 \mu_0 \Lambda$, since their supercurrents and magnetic fields essentially cancel everywhere.

### D. Work done by the external current supply when a vortex appears

If a vortex appears at a distance $\rho$ from the origin, the flux through the drive coil at $z = -R/\sqrt{2}$ changes by: $\Delta \phi = \pi a^2 B_{V,z}(\rho, z = -R/\sqrt{2})$. In the strong-screening limit we have: $\Delta \phi \approx a^2 \Phi_0 R / (2\rho^2 + R^2)^{3/2}$, where "$a$" is the radius of the drive coil. The net work done to keep the current in the drive coil constant is: $I_d \Delta \phi$, where $I_d = B_0 R^3 / \sqrt{2} \mu_0 a^2$ is the current in the drive coil necessary to produce field $B_0$. Thus:

$$W(\rho) = \frac{B_0 \Phi_0 R}{\sqrt{2} \mu_0} \frac{1}{(1 + 2\rho^2/R^2)^{3/2}} \ . \qquad (\Lambda/R \ll 1) \tag{16}$$

We see that $-W(\rho)$ is a potential well that attracts vortices and repels antivortices, so that when a V-aV pair unbinds, the vortex moves toward the origin, and the antivortex moves away until the long-range V-aV attraction stops it.

### E. Critical external perpendicular field $\mathbf{B}_{c1}^0$

The defining condition for the external critical field is that the work done by the external current supply when the first vortex and antivortex appear, Eq. (16), equals the Helmholtz free energy of the vortex and antivortex. We therefore define the Gibbs free energy $\Delta G$ as the extra Helmholtz free energy of a vortex at $\boldsymbol{\rho}_V$ and an antivortex at $\boldsymbol{\rho}_{aV}$, minus the work done in their creation. Assuming both vortices lie on the same ray from the origin, and defining: $\rho_{12} \equiv \rho_{aV} - \rho_V$, we can write:

$$\frac{\Delta G(B_0, \rho_V, \rho_{aV})}{2U_V} = 1 - \frac{1}{\ln(\Lambda/\xi)} \ln\left(1 + \frac{\Lambda}{\rho_{12}}\right)$$

$$- \frac{B_0}{B_{c1}^0} \left[ \frac{1}{\left(1 + 2\rho_V^2/R^2\right)^{3/2}} - \frac{1}{\left(1 + 2\rho_{aV}^2/R^2\right)^{3/2}} \right], \quad (\Lambda/R \ll 1) \quad (17)$$

where $B_{c1}^0$ is:

$$B_{c1}^0 = \frac{4\sqrt{2}\Lambda}{R} \frac{\Phi_0}{4\pi\Lambda^2} \ln\left(\frac{\Lambda}{\xi}\right). \quad (18)$$

Equation (17) applies in the strong-screening limit. Note that $\frac{\Phi_0}{4\pi\Lambda^2} \ln\left(\frac{\Lambda}{\xi}\right)$ in Eq. (18) is the 2-D intrinsic critical field, $B_{c1}^{2D}$, defined as in 3-D, but with $\lambda$ replaced by $\Lambda$.

It is easily seen from Eq. (17) that $\Delta G = 0$ when: 1) $B_0 = B_{c1}^0$; 2) the vortex is at the origin; and 3) the antivortex is at infinity. Because the V-aV interaction is long-ranged, the actual equilibrium position, $\rho_{aV}^{eq}$, of the antivortex is not at infinity but rather at: $\rho_{aV}^{eq} \approx R\left[\frac{R}{\Lambda} \ln\left(\frac{\Lambda}{\xi}\right)\right]^{1/2}$, which means $\rho_{aV}^{eq} \approx 10R$ for typical sample parameters. This means that $\Delta G(B_{c1}^0, 0, \rho_{aV})$ is actually slightly negative and a minimum at $\rho_{aV} \approx \rho_{aV}^{eq}$, and the external critical field is therefore a tiny bit smaller than $B_{c1}^0$ given in Eq. 18. The prefactor, $4\sqrt{2}\Lambda/R$, in Eq. (18) captures the effect of demagnetization.[11]

From the foregoing, we see that a film's radius is effectively infinite only if it is much larger than both $\Lambda$ and $\rho_{aV}^{eq}$. Also, we note that there is entropy associated with the angular position of the antivortex, $S_{aV} \approx k_B \ln(2\pi\rho_{aV}^{eq}/\xi)$, and we have neglected its contribution, $-TS_{aV}$, to $\Delta G$. As $B_0$ increases beyond $B_{c1}^0$, vortices accumulate near the film center, and the belt of corresponding antivortices moves closer in, e.g., when the equilibrium state has six vortices, the six antivortices are only half as far away as the first antivortex.

The critical field that we calculate for the conventional geometry of a circular film in a uniform external perpendicular field agrees well with Fetter's.[10] In this geometry the first vortex-bearing state has a single vortex at the origin. Since the applied field is uniform, the work done when that vortex appears can be calculated from:[15] $W = B_0 \int dA \, tM_z(\rho)$, where the magnetization is: $tM_z(\rho) \equiv \rho \times K_V(\rho)/2$. $\rho$ is a 2-D displacement vector, and $K_V(\rho)$ is the vortex sheet current density. Using a numerical calculation of $K_V(\rho)$ that captures the increase in $K_V(\rho)$ near the film perimeter, we find that the work done by the current supply is: $W \approx 1.25 B_0 \Phi_0 R_f / \mu_0$, and therefore the external critical field is:

$$B_{c1}^0 \approx \frac{1.6\Lambda}{R_f} \frac{\Phi_0}{4\pi\Lambda^2} \ln\left(\frac{\Lambda}{\xi}\right). \qquad \text{(uniform external field)} \qquad (19)$$

This field is only about 60% higher than Fetter's[10] numerical result in the large-radius, strongly screening film limit (see Fig. 5 and associated text in Ref. 10), the difference being due to different treatments of the increase in $K_V(\rho)$ near the film perimeter.

In seeming contradiction to our result, Mawatari and Clem[12] (MC) find that vortices first enter a large-radius, "thick", film ($\xi \ll t \ll \lambda$) in a non-uniform applied field (produced by a straight wire parallel to the film) when the maximum parallel field at the sample surface reaches: $B_{c1}^{\|} \approx \frac{2\Phi_0}{\pi t^2} \ln\left(\frac{2t}{\pi\xi}\right)$,[16-18] for a strongly screening film without vortex pinning. For comparison

with the present paper, MC's conclusion can be rephrased as: vortices first appear when the maximum applied perpendicular field is: $B_0 = \frac{\Phi_0}{2\pi t^2} ln\left(\frac{2t}{\pi\xi}\right)$. For quasi-2D films, i.e., a few coherence lengths thick, this is orders of magnitude larger than our external critical field. We propose that MC's critical field is the field at which the barrier for creating vortex-antivortex pairs vanishes.

### III. SUMMARY

Motivated by a desire to understand nonlinear effects in two-coil measurements, we calculate the external lower critical field $B_{c1}^0$ when a non-uniform magnetic field is applied to an infinite-radius thin superconducting film. The first vortex-bearing state has both a vortex and an antivortex, the former near the origin and the latter far from the origin, but not at infinity due to the long-range V-aV attraction. The effective external force acting on vortices comes from the work done by the drive-coil's current supply when vortices move. The radial distance, $R$, where the non-uniform field changes sign emerges in the same role that the film radius, $R_f$, plays in the uniform-field configuration. In the lab, strong nonlinearities usually appear at applied fields much larger than calculated here, indicating that the appearance of vortices is inhibited by vortex pinning and/or the free energy barrier for breaking nascent V-aV pairs.

### Acknowledgements

We are grateful to John Clem, Yen Lee Loh, and Alex Gurevich for insightful comments. This work was supported in part by DOE-Basic Energy Sciences through Grant No. FG02-08ER46533, and in part by NSF grant DMR-0805227.